# Development of HPLC-Orbitrap method for identification of N-bearing molecules in complex organic material relevant to planetary environments

*Thomas Gautier*[*1,2], *Isabelle Schmitz-Afonso*[3,4], *David Touboul*[3], *Cyril Szopa*[2,5], *Arnaud Buch*[6], *Nathalie Carrasco*[2,5]

[1] NASA Goddard Space Flight Center, code 699, 8800 Greenbelt Rd, Greenbelt, MD 20771, USA

[2] LATMOS/IPSL, UVSQ Université Paris-Saclay, UPMC Univ. Paris 06, CNRS, Guyancourt, France

[3] Institut de Chimie des Substances Naturelles ICSN-CNRS UPR 2301, Université Paris-Sud, 1 avenue de la terrasse, 91198, Gif-sur-Yvette, France

[4] Normandie Université, COBRA, UMR 6014 et FR3038, Université de Rouen ; INSA de Rouen ; CNRS, IRCOF, 1 rue Tesnière, 76821 Mont-Saint-Aignan Cedex, France

[5] Institut Universitaire de France, 103 Bvd St-Michel, 75005 Paris, France

[6] LGPM, Ecole Centrale de Paris, Grande voie des Vignes, 92295 Chatenay-Malabry Cedex, France

Corresponding author, current address:

*Thomas Gautier
NASA – Goddard Space Flight Center
8800 Greenbelt Road
Greenbelt, MD 20771
Phone:
Mail: thomas.j.gautier@nasa.gov

**Highlights**

➤ We analyze Titan's tholins using HPLC-Orbitrap

➤ We strictly identify the isomers of seven of the major molecules constituting tholins

➤ All confirmed molecules bear nitrogen and most of them are aromatics.

➤ This supports the hypothesis of a tholins formation passing through PANH

## Abstract

Although the Cassini Spacecraft and the Huygens lander provided numerous information about Titan atmospheric chemistry and the formation of its aerosols, the exact composition of these aerosols still remains unknown. A fruitful proxy to investigate these aerosols is the use of laboratory experiments that allow producing and studying analogs of Titan aerosol, the so-called tholins. Even when produced in the laboratory, unveiling the exact composition of the aerosol remains problematic due to the high complexity of the material. Numerous advances have been recently made using high-resolution mass spectrometry (HRMS) (Pernot *et al.* 2010, Somogyi *et al.* 2012, Gautier *et al.* 2014) that allowed the separation of isobaric compounds and a robust identification of chemical species composing tholins regarding their molecular formulae. Nevertheless isomeric species cannot be resolved by a simple mass measurement. We propose here an analysis of tholins by high performance liquid chromatography (HPLC) coupled to HRMS to unveil this isomeric ambiguity for some of the major tholins compounds. By comparing chromatograms obtained when analyzing tholins and chemical standards, we strictly identified seven molecules in our tholins samples: melamine, cyanoguanidine, 6-methyl-1,3,5-triazine-2,4-diamine, 2,4,6-triaminopyrimidine, 3-amino-1,2,4-triazole, 3,5-Dimethyl-1,2,4-triazole and 2,4-diamino-1,3,5-triazine. Several molecules, including hexamethylenetriamine (HMT) were not present at detectable levels in our sample. The use for the first time of a coupled HPLC-HRMS technique applied to tholins study demonstrated the interest of such a technique compared to single high-resolution mass spectrometry for the study of tholins composition.

# 1. Introduction

The atmosphere of Titan is mainly constituted of $N_2$ and $CH_4$. Solar irradiation and electrically charged particles accelerated in Saturn's magnetosphere (Sittler Jr. *et al*. 2009) induce organic chemical reactions within the atmosphere. These reactions lead to the production in Titan's atmosphere of an opaque layer of organic solid aerosol. To study these Titan's aerosols, it is possible to produce laboratory analogs, the so-called "tholins". A discussion on the different experimental setups designed for such a purpose can be found in Cable et al. (2012). The properties of the produced tholins allow a better analysis and understanding of observational data obtained in the atmosphere of Titan.

Recent advances in HRMS, with Orbitrap$^{TM}$ or Fourier-Transform Ion Cyclotron Resonance mass spectrometers, gave a first proxy on the aerosol constitution and revealed that tholins are formed of thousands of different chemical compounds (Pernot *et al*. 2010, Somogyi *et al*. 2012, Gautier *et al*. 2014).

However, each detected mass peak only gives access to the raw molecular formula of the compound whereas, especially for high masses, one formula can be attributed to several different isomeric molecules bearing highly different structures and reactivity. It is then of prime importance to be able to strictly identify the molecules within the tholins, by separating all the isomers. The separation methods mostly used for tholins analysis are pyrolysis GC/MS (Khare *et al*. 1984, McGuigan *et al*. 2006, Coll *et al*. 2013) and GC/MS (Pilling *et al*. 2009, He and Smith 2014). Thin-layer chromatography prior to mass spectrometry (Jagota *et al*. 2014) and microelectrophoresis (Cable *et al*. 2014) were recently used whereas very few liquid chromatography separations of tholins have been published (Ruiz-Bermejo *et al*. 2008, Cleaves *et al*. 2014). As mentioned by Cable *et al*. 2012, liquid chromatography could provide a more comprehensive analysis of the tholins molecular structure with a separation for both polar and charged species.

Therefore, to go further in tholinomics and unveil the degeneracy of the molecular formulae, we separated the tholins soluble fraction by high-performance liquid chromatography (HPLC) coupled to a hybrid linear trap/Orbitrap$^{TM}$ mass spectrometer and compared retention times with those of standard compounds. The method presented in this paper has been developed for Titan's tholins but it could be applied to any complex organic material in the solar system such as meteorite organic soluble matter, kerogens, interstellar ice analogs etc.

## 2. Experimental

### 2.1. Tholins preparation and HPLC-HRMS developments

Tholins were produced with a cold plasma in a 95-5% $N_2$-$CH_4$ mixture using the PAMPRE reactor (Szopa *et al*. 2006). They were produced and prepared following the procedure described in Gautier *et al*. (2014). These tholins were previously analyzed by direct HR-MS in Gautier *et al*. 2014.

For mass spectrometric analysis, tholins were first dissolved in methanol (HPLC grade, Baker) at a concentration of 4 mg.mL$^{-1}$, then mixed thoroughly. The resulting solution was filtered through a PTFE 0.2 µm membrane to remove remaining tholins that did not dissolve in methanol (Carrasco et al. 2009). HPLC was performed with a HPLC Ultimate 3000 system (Dionex). Three different columns were tested to achieve proper and complementary separation. First we used an octadecyl C18 column 2.1 mm internal diameter (i.d.), 100 mm, particle size 3.5 µm (Zorbax SB-Aq, Agilent technologies), equipped with a guard column and thermostated at 25°C. Elution was performed with a mobile phase of pH=6.85 consisting of 20 mM of ammonium acetate in water (A) and acetonitrile (HPLC grade, Baker) (B) with the following gradient: 0 to 2 min, isocratic with 0% B – 2 to 11 min, 0 to 21% B - 11 to 13 min, 21 to 100% B then maintaining 3 min at 100% B to rinse the column and finally return to initial conditions and stabilization of the column. Second, a cyano column 2.1 mm i.d., 150 mm, particle size 5 μm (Zorbax Eclipse XDB-CN, Agilent technologies), thermostated at 25°C, was tested. Elution was performed with a mobile phase consisting of 20 mM of ammonium acetate in water (A) and acetonitrile (B) with the following gradient: 0 to 3 min, isocratic with 4% B – 3 to 15 min, 4 to 100% B then maintaining 3 min at 100% B to rinse the column and finally return to initial conditions and stabilization of the column. The third column was an aminopropyl column 2 mm i.d., 150 mm, particle size 3 μm (Luna $NH_2$, Phenomenex), thermostated at 20°C. This column was operated in hydrophilic interaction chromatography mode with a mobile phase consisting of 20 mM of ammonium acetate in water (A) and acetonitrile (B): 0 to 10 min, 95 to 73% B – 10 to 17 min, 73 to 50% B then maintaining 8 min at 50% B to rinse the column and finally return to initial conditions and stabilization of the column. The tholins extract was diluted by a factor of two to get a sample in water/methanol (50/50), for the reversed phase separations (octadecyl and cyano columns) and in acetonitrile/methanol (50/50) for the hydrophilic interaction chromatography (aminopropyl column). For the three columns, flow rate was set to 0.25 mL.min$^{-1}$ and injection volume was 7 μL.

The HPLC was directly coupled to a hybrid linear trap/Orbitrap$^{TM}$ (LTQ/Orbitrap$^{TM}$, ThermoScientific) mass spectrometer equipped with an ElectroSpray Ionization (ESI) source. Analyses were performed in the positive ion mode. Acquisition parameters of the ESI-LTQ/Orbitrap$^{TM}$ were: needle voltage 4.5 kV; capillary temperature 275°C; capillary voltage 25 V; tube lens voltage 65 V; sheath gas flow rate 40 arbitrary unit (a.u.). The mass spectrometer was externally calibrated using caffeine, MRFA (met-arg-phe-ala) peptide and Ultramark 1621 allowing a mass precision below 2 ppm. Data were acquired with a mass resolution set to 100 000 at *m/z* 450. The experimental resolution m/Δm was determined to be ~200 000 at *m/z* 150 for both MS and MS/MS measurements.

## 2.2 Molecules of interest; choice of standards

In the present work we aim to determine the exact composition of tholins produced with PAMPRE (introducing 5% of methane in dinitrogen for the gas mixture) by determining the exact composition (formula and structure) of the most abundant peaks detected in the high-resolution mass spectrum presented in Figure 1 (Gautier *et al*. 2014). Samples analyzed in the present work were produced following the same protocol as the one described in Gautier et al. 2014.

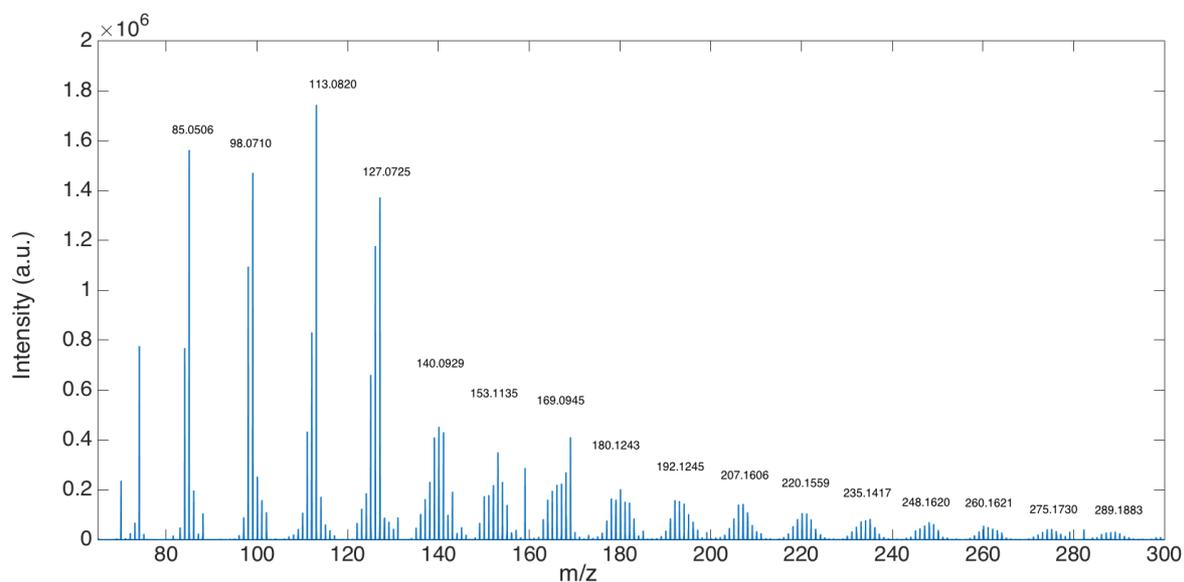

**Figure 1: Mass spectrum of an extract of tholins (infusion of the sample) produced with 5% of methane in nitrogen. Numbers in the figure indicate the most intense ion for each cluster. Adapted from Gautier et al. 2014**

The interest of the method proposed here is to unveil the isomeric ambiguity remaining after high-resolution Orbitrap analysis and exact mass determination. However, the high complexity of the mass spectrum of tholins obtained with the Orbitrap (>15,000 peaks detected) makes an exhaustive analysis of all peaks impossible. We focused here on a limited set of compounds to be tested with our method.

The compounds of interest for HPLC-MS analysis are chosen based on two criteria. As a first approach, we limit this study to the most intense ions detected in this spectrum that might represent a significant portion of the material composing sample (see table 1). Second, we choose the standard compounds depending on their availability and affordability for comparison of retention time. The complete set of compounds tested can be find in Table 1 for the compounds detected in tholins and in Table SI 1 for those non confirmed to be present in tholins.

Standards 1, 2, 5, 8, 9 and 10 were purchased from VWR. Standards 3, 4, 6, 7, 11 and 12 were purchased from Sigma Aldrich. Standard 13 was purchased from Fluka. All stock solutions were prepared in water except for standards 7, 10, 12 and 13 prepared in MeOH/water, 50/50, v/v to be able to solubilize them. After dilution in methanol/water or methanol/acetonitrile, each individual standard and a mixture of all standards were analyzed in the three chromatographic conditions developed with the tholins extract. Validation of the presence of the standard in the tholins extract was done by retention time comparison. Concentrations tested ranged from $3.10^{-4}$ to $1.10^{-2}$ mg.mL$^{-1}$, varying for each compound.

## 3. Results

The three columns, octadedyl, cyano and aminopropyl, have been implemented for complementary chromatographic separation in regard with known tholins chemical properties. The octadecyl column, with an apolar stationary phase, has been chosen for reversed phase separation starting with 100% aqueous phase to get maximum retention of polar analytes, with retention time increasing with decreasing polarity of the analyzed compounds. The cyano column with a more polar stationary phase, dimethyl-cyanopropylsilane, is especially suited for separation of highly polar, acidic and basic compounds with better retention of polar analytes through interactions with the stationary phase in reversed phase mode. The aminopropyl column has been selected to work with hydrophilic interaction chromatography (HILIC), thus inducing different retention capabilities (Buszewski and Noga 2012). Two types of interaction may occur in HILIC mode in our study: hydrophilic interactions with retention increasing with polarity, and electrostatic repulsion between positively charged analytes and the positively charged stationary phase. In the particular case of our basic compounds, no hydrogen bonding may occur with the stationary phase. Furthermore, HILIC is particularly adapted for polar nitrogen compounds, which are one of the major parts of the tholins composition (Zheng *et al* 2012).

### 3.1. Interest of the method

Figure 2 provides the Base Peak Chromatogram (BPC) of tholins sample using the aminopropyl column. This congested BPC of tholins denotes the need of both separation with HPLC and high resolution in mass with Orbitrap$^{TM}$ to isolate precisely compounds from the bulk material in order to identify them.

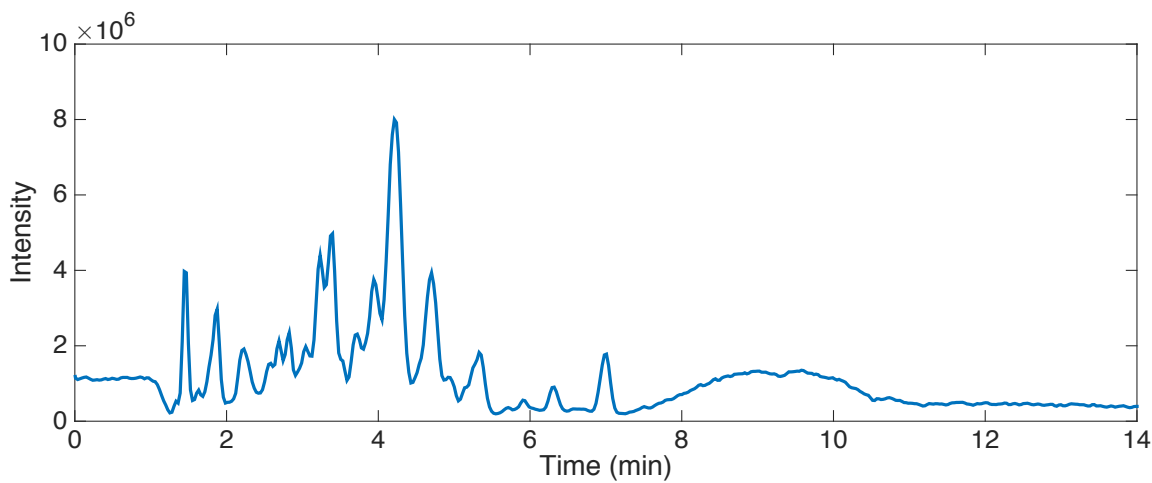

**Figure 2: Base Peak Chromatogram (BPC) Intensity chromatogram of a tholins sample analyzed with the aminopropyl column**

The complete set of chromatograms is given in the supplementary information together with an example of blank experiment (figure SI7). We provide here an example set of chromatograms at *m/z* 85 and 98 in the tholins extract and in the mixture of standards, with separation by aminopropyl, cyano and octadecyl columns.

Figure 3 shows the different extracted ion chromatograms (EIC) for *m/z* 85. Figure 3a represents the elution using an aminopropyl column, 3b using a cyano column and 2c using a C18 column. Figure 4 represents the same set of data for *m/z* 98. These two figures clearly show the complementarity of the different stationary phases for the separation of the different analytes presents in the tholins. Figure 3 demonstrates that the aminopropyl column leads to the best separation of isomers at *m/z* 85 present in tholins, as well as the two standards. On figure 4, standards 5 and 6 are better separated using the C18 column. Standard 5 retention time does not match with any tholins compound with the C18 column whereas it seems to match with a tholins compound with the amino and cyano columns. Standard 6 is confirmed with the three columns. The aminopropyl column evidences many isomeric compounds for the same exact mass. Regarding the eight *m/z* ratio studied here (Other EIC are presented in the Supplementary Information), the most efficient column is the aminopropyl column with around 39 different isomers separated, then the octadecyl column with 25 compounds and finally the cyano column with 20 compounds. Only major compounds with a signal-to-noise ratio above 100 (Xcalibur Thermo software, automatic integration) were taken into account for each EIC. The aminopropyl stationary phase in hydrophilic chromatography seems the most powerful for the separation of tholins isomers but octadecyl phase in reversed phase mode stays useful for complementary separation for some isomers.

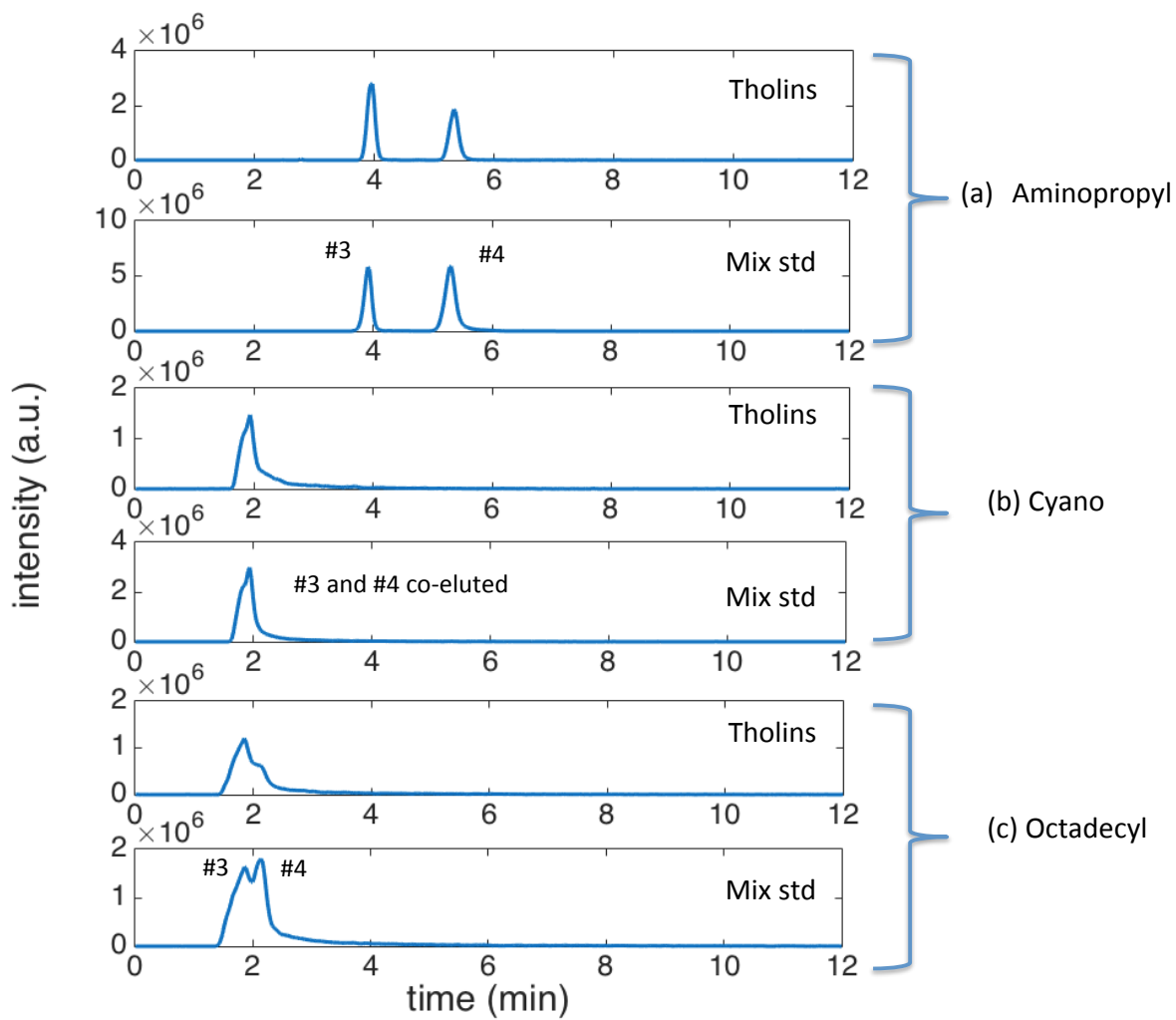

Figure 3: Extracted ion chromatograms of *m/z* 85.0500 to 85.0518, tholins sample and standards 3 and 4, for columns (a) aminopropyl (b) cyano (c) octadecyl.

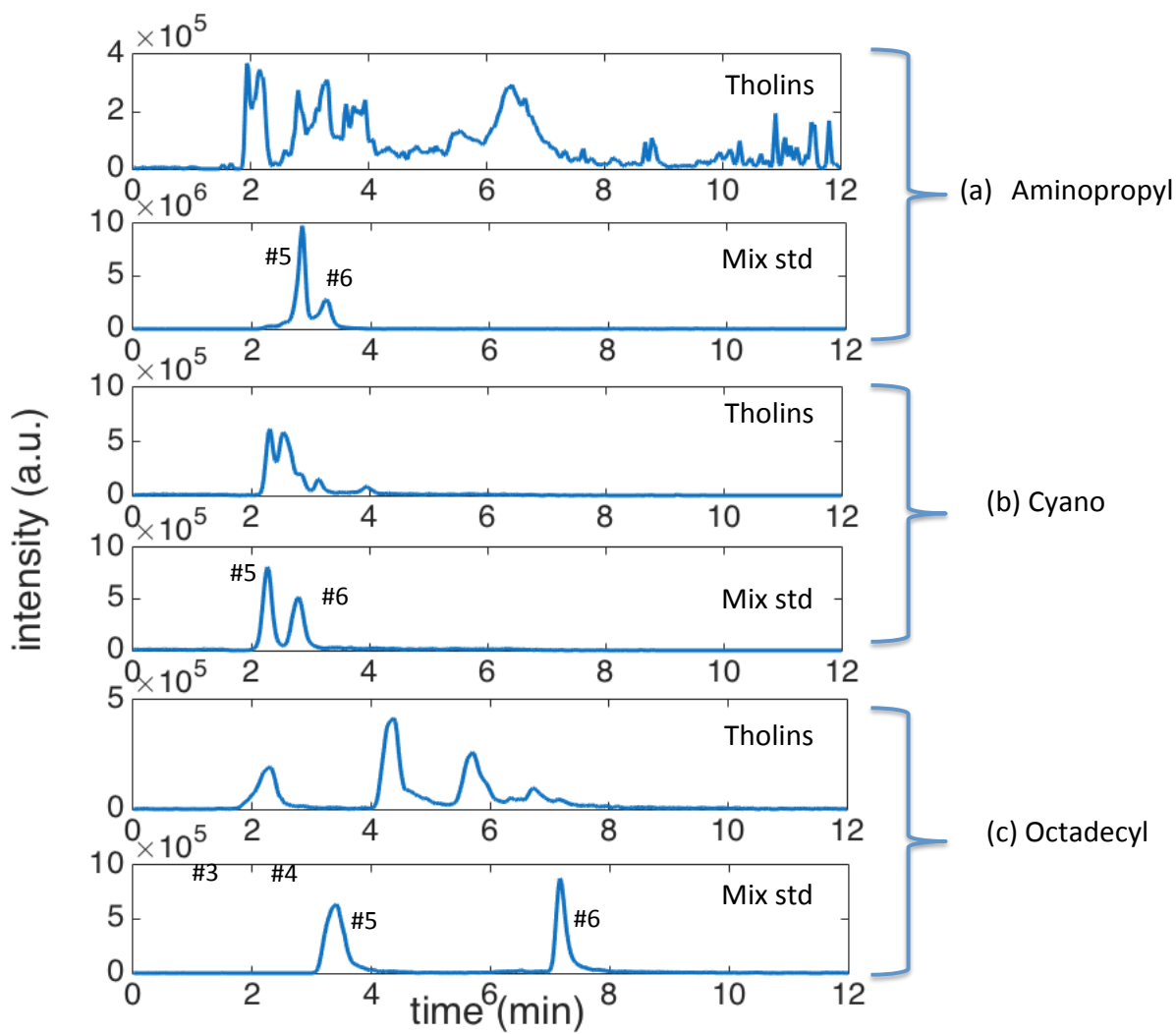

Figure 4: Extracted ion chromatograms of *m/z* 98.0706 to 98.0720, tholins sample and standards 5 and 6, for columns (a) aminopropyl (b) cyano (c) octadecyl.

### 3.2. Standard confirmation

The retention time obtained both for the tholins sample and the standard tested are summarized in table 1. A complete set of chromatograms is given in supplementary material.

**Table 1:** Column one to six respectively give the *m/z* ratio, the standard number, the name, the structure, the CAS Number and the Formulae of standard compounds tested and confirmed to be present in tholins. Retention time of sample and confirmed standard on the three columns are given columns seven to twelve. Values correspond to the average on three injections and displayed uncertainties were calculated at 1 σ. A list of retention time for compounds not confirmed to be present in tholins can be find in Table SI 1 of the supplementary material.

| [M+H]+ m/z | Std | Name | Structure | CAS Number | Formulae | Aminopropyl | | Cyano | | Octadecyl | |
|---|---|---|---|---|---|---|---|---|---|---|---|
| | | | | | | Rt std (min) | Rt tholins (min) | Rt std (min) | Rt tholins (min) | Rt std (min) | Rt tholins (min) |
| 85.0509 | 4 | Cyanoguanidine | 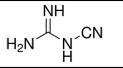 | 461-58-5 | C2N4H5 | 3.97 ± 0.08 | 3.95 ± 0.02 | 1.93 ±0.02 | 1.93 ±0.00 | 1.85 ±0.02 | 1.85 ± 0.00 |
| | 3 | 3-amino-1,2,4-triazole | 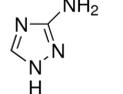 | 61-82-5 | C2N4H5 | 5.31 ±0.00 | 5.33 ± 0.00 | 1.94 ±0.02 | 1.93 ±0.00 | 2.14 ±0.01 | 2.13 ± 0.00 |
| 98.0713 | 5 | 5-methyl-3-amino-1,2-pyrazole | 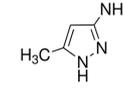 | 31230-17-8 | C4N3H8 | 2.85 ±0.01 | 2.81 ± 0.02 | 2.27 ±0.00 | 2.32 ±0.03 | 3.41 ±0.00 | 4.36 ± 0.02 |
| | 6 | 3,5-Dimethyl-1,2,4-triazole | 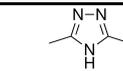 | 7343-34-2 | C4N3H8 | 3.26 ±0.00 | 3.26 ± 0.03 | 2.79 ±0.02 | 2.82 ±0.00 | 7.17 ±0.02 | 7.15 ±0.03 |
| 112.0618 | 7 | 2,4-diamino-1,3,5-triazine | 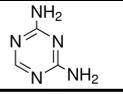 | 504-08-5 | C3N5H6 | 4.16 ±0.00 | 4.16 ± 0.00 | 2.31 ±0.01 | 2.29 ±0.00 | 6.65 ±0.00 | 6.67 ± 0.01 |
| 126.0774 | 8 | 6-methyl-1,3,5-triazine-2,4-diamine | 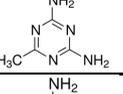 | 542-02-9 | C4N5H8 | 4.21 ±0.02 | 4.22 ± 0.02 | 2.77 ±0.00 | 2.77 ±0.00 | 8.04 ±0.02 | 8.04 ± 0.02 |
| | 9 | 2,4,6-triaminopyrimidine | 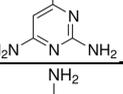 | 1004-38-2 | C4N5H8 | 6.28 ±0.00 | 6.31 ± 0.02 | 3.22 ±0.03 | 3.24 ±0.00 | 6.29 ±0.02 | 6.31 ± 0.02 |
| 127.0727 | 10 | Melamine | 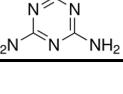 | 108-78-1 | C3N6H7 | 6.97 ±0.02 | 6.99 ± 0.02 | 2.19 ±0.00 | 2.21 ±0.00 | 6.08 ±0.00 | 6.10 ± 0.00 |

1  Using the HPLC coupling with high resolution MS we have been able to identify strictly the
2  following molecules in our tholins sample: 3-amino-1,2,4-triazole, cyanoguanidine, 3,5-

Dimethyl-1,2,4-triazole, 2,4-diamino-1,3,5-triazine, 6-methyl-1,3,5-triazine-2,4-diamine, 2,4,6-triaminopyrimidine and melamine (standards 3, 4, 6 to 10 respectively). To our knowledge, and with exception of Cyanoguanidine and Melamine, none of these molecules have been previously reported in the literature to be present in tholins. These results highlighted the importance of separation methods coupled to mass spectrometry to characterize the compounds present in tholins as they present many isomeric species. In direct introduction methods, even with a high-resolution instrument, these isomeric compounds cannot be differentiated and MS/MS spectra would be representative of several compounds thus inducing interpretation errors. Our results indicate that 1-methyl-1,2,4-triazole, 3-aminopyrazole, (S)-5-(2-pyrrolidinyl)-1H-tetrazole, 4-amino-2-dimethylamino-1,3,5-triazine and HMT (standards 1, 2, 11, 12 and 13, respectively) are not detected in our tholins sample with the analytical method set up. For standard 5, the result is more ambiguous since there are many isomers detected in the tholins sample; this standard would be confirmed only on the aminopropyl and cyano column. As we consider that the three columns should validate each compound, we would conclude that 3-amino-5-methylpyrazole (standard 5) is not detected in PAMPRE tholins.

To provide a rough estimate of compounds concentration in tholins soluble fraction we used the standard addition method. We measured the area of a given compound in the tholins chromatogram ($A_t$) and the area of the same peak after adding a known amount ($C_s$) of standard compound in the tholins sample ($A_{ts}$). Concentration of this compound in tholins solution can therefore be estimated as $C_t=(A_t*C_s)/(A_{ts}-A_t)$. Concentrations for compounds detected in tholins are given in table 2. We would like to emphasize though that these values represent only a rough estimate of compound concentration in tholins solution. Given the fact that tholins are not 100% soluble (Carrasco et al. 2009) and that solubility of tholins might differ from one compound to another, it is impossible to interpret these values as actual concentration within the bulk tholins. However, values provided in table 2 can be used as lower limits for compounds concentration in tholins.

**Table 2: Calculation of identified compounds concentration in tholins soluble fraction ($C_t$). $C_s$ represents the known concentration of standard added, $A_{ts}$ the area of the chromatographic peak for tholins + standard (averaged on 3 injections), and $A_t$ the area of the peak for tholins solely (averaged on 3 injections).**

| $[M+H]^+$ $m/z$ | Std number | $C_s$ (mg.mL$^{-1}$) | $A_{ts}$ | $A_t$ | $C_t$ (mg.mL$^{-1}$) |
|---|---|---|---|---|---|
| 85.0509 | 4 | 5.00E-03 | 5.32E+07 | 2.72E+07 | 5.2E-03 |
|  | 3 | 2.50E-03 | 7.36E+07 | 2.07E+07 | 9.8E-04 |
| 98.0713 | 6 | 2.50E-04 | 3.07E+07 | 5.00E+06 | 4.9E-05 |
| 112.0618 | 7 | 5.00E-03 | 1.86E+08 | 6.07E+07 | 2.4E-03 |
| 126.0774 | 8 | 2.50E-04 | 1.27E+08 | 1.03E+08 | 1.1E-03 |
|  | 9 | 5.00E-04 | 2.89E+07 | 9.17E+06 | 2.3E-04 |
| 127.0727 | 10 | 1.25E-03 | 3.76E+07 | 1.88E+07 | 1.3E-03 |

A compound of interest we were particularly looking for in our sample was HMT (standard 13). HMT is a molecule detected in the final products of many astrophysical laboratory experiments, especially the one simulating interstellar ice irradiations (Bernstein *et al.* 1995, Muñoz Caro *et al.* 2003, Vinogradoff *et al.* 2012). The detection of HMT in laboratory tholins *a priori* similar to our samples has been reported recently by He *et al.* (2012) using NMR and high resolution mass spectrometry and was proposed to be a major component of their tholins.

In our case a major product at *m/z* 141.1131 was detected corresponding to the exact mass ($m/z_{th.}$ 141.1135; $\Delta_{m/z}$ = -2.4 ppm) and thus formulae of protonated HMT. However our HPLC results show that no signal at *m/z* 141 was detected in the tholins sample at the retention time of the HMT standard (Figure SI6). Furthermore, MS/MS spectra of the isomers in tholins sample (Figure 5a) and HMT (Figure 5b) show different fragmentation patterns. Fragmentation pattern also does not match any HMT derivative detected in ice irradiation (Danger *et al.* 2013). For the tholins sample, the losses of $NH_3$ and HCN, already described in MS/MS experiments of tholins (Somogyi *et al.* 2005), would suggest either the presence of $NH_2$ function and/or a cyano function and/or a cycle including a nitrogen atom.

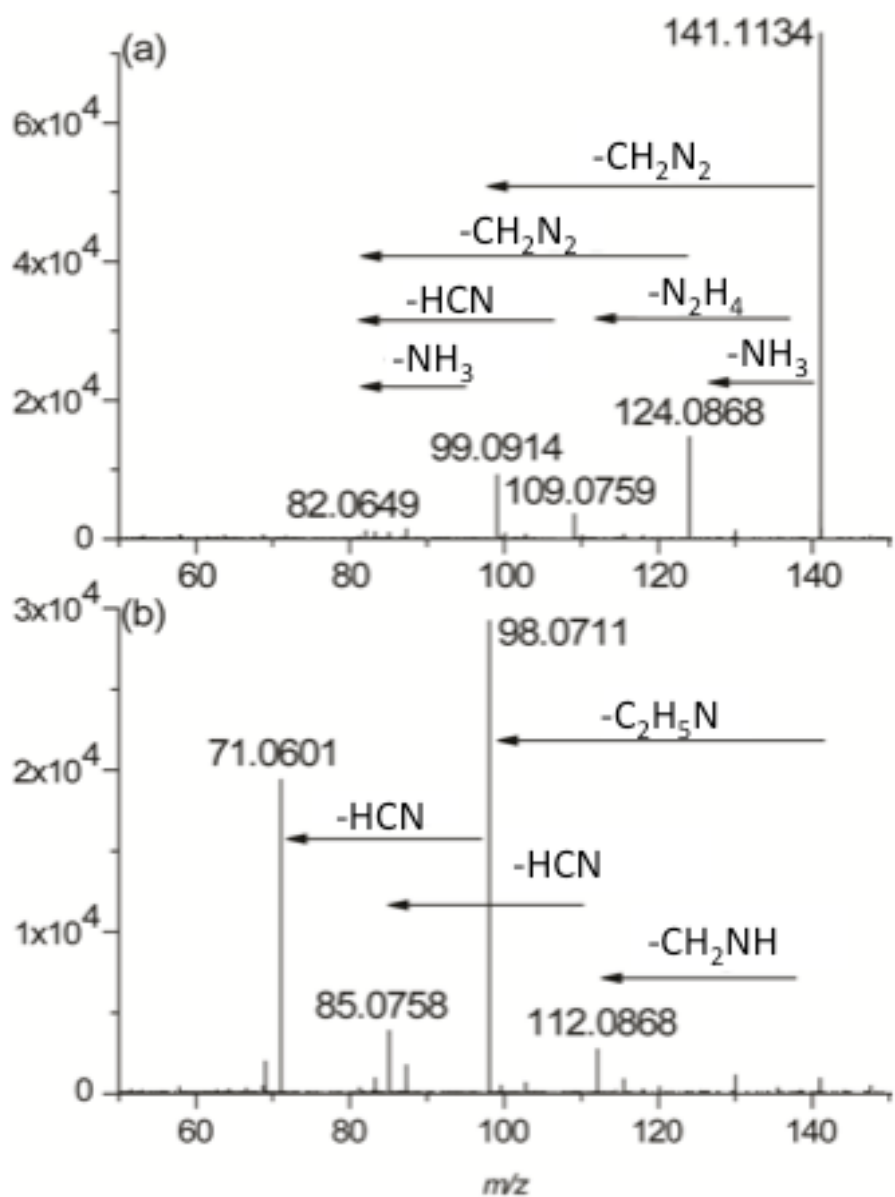

Figure 5: LC-MS/MS spectra of *m/z* 141.1131 ($C_6H_{13}N_4$) for (a) tholins isomer (b) HMT, recorded with the same collision energy, CE = 24 a.u., activation $q_z$ 0.25.

53  This clearly indicates that, at least in our sample, the major compound detected at *m/z*
54  141.1131 cannot be attributed to HMT. The isomeric compound detected in our tholins
55  sample is less polar than HMT regarding its higher retention time for octadecyl column and
56  shorter retention time for aminopropyl column.

57  ## 4. Discussion

58  This difference between our results and the one from He *et al*. (2012) is difficult to
59  understand as the pathway they proposed for HMT formation passes through gas phase
60  reactions of methanimine. Methanimine has been firmly detected as a volatile compound
61  present in the PAMPRE reactor during standard experiments (Carrasco *et al*. 2012), that
62  would lead to the formation of HMT according to He *et al*. proposed mechanism.

63  Considering the complexity of the molecular identification in tholins samples (which required
64  comparisons to standards in our case), a first explanation for the discrepancies between the
65  experiments might then be that the compound identified by He *et al*. (2012) using NMR is not
66  HMT but one of its isomer, as in PAMPRE tholins.

67  Another possibility would be that HMT is actually produced in the case of He *et al* (2012);
68  but with a mechanism specific to this experiment. Indeed He *et al.* 2012 used an AC plasma,
69  where the solid samples remains in the reactor for the whole duration of the synthesis whereas
70  the tholins in PAMPRE are dragged out of the plasma after a few minutes. The solid samples
71  are therefore longer exposed to the plasma bombardment in the case of He *et al* (2012).
72  Moreover in the case of He *et al*. 2012, tholins are not collected directly after production at
73  cold temperature but are warmed up to room temperature for 24 h before collection. Recent
74  investigation on PAMPRE showed that such a process tends to form a complex organic
75  residue in addition to the tholins by reaction taking place within the solid state (Gautier *et al*.
76  2014). These secondary reactions may explain the formation of HMT, whose formation pathways are allowed by thermal reactivity in cometary ice analogs (Vinogradoff *et al*. 2011, Vinogradoff *et al*. 2012, Theulé *et al*. 2013).

Among the compounds detected in our sample are cyanoguanidine (standard 4) and melamine (standard 10). HPLC-LTQ/Orbitrap™ provides here a firm detection of these two compounds recently suggested to be present in tholins by NMR studies, in sample produced both with PAMPRE (Derenne *et al*. 2012) and with another experimental setup (He and Smith 2013, He and Smith 2014).

As stated by He *et al*. 2014, cyanoguanidine and melamine are a dimer and a trimer of cyanamide, respectively. These two compounds may be of interest regarding aerosols formation in Titan's atmosphere. In particular, melamine is well known for its polymeric properties and tends to form large-scale structures of highly cross-linked material (Wohnsiedler 1953, Philbrook *et al*. 2005, Merline *et al*. 2013) that could serve as a base structure for aerosol growth.

Regarding the molecules present in our tholins, all have high nitrogen content and six out of seven are aromatic molecules. As a reminder these molecules correspond to the most predominant peaks in the mass spectrum of tholins recorded with an Orbitrap™ analyser. Even though N-bearing compounds do not form the entire material (some pure hydrocarbons are detected by Orbitrap™ analyser, see Pernot *et al*. (2010), Gautier *et al*. (2014)), it is clear that N-bearing aromatics constitute a significant part of tholins material. Triazine-like cycle is a predominant feature in the compounds detected in this study. This is in agreement with results from Quirico *et al*. (2008) that also proposed triazine as an important contribution to tholins signature in infrared absorption spectroscopy. The predominance of such compounds in tholins is a possible indication of N-Bearing Polyaromatic Hydrocarbons (PANH) formation pathway, rather than a pure Polyaromatic Hydrocarbons (PAH) one.

## 5. Conclusion

We used a complementary HPLC-Orbitrap$^{TM}$ HRMS analysis on Titan's tholins and standard compounds. Our study confirms the importance of a proper chromatographic separation for the analysis of complex (several thousands of peaks in the MS) organic mixtures, such as tholins, where a single high resolution MS peak can be due to the contribution of a dozen of different isomers. We saw that in such a case two chromatographic columns, at the very least, are necessary for a proper decomposition of the material due to the large variability in polarity of the compounds composing tholins. In case of using only two columns we would recommend the C18 in reversed phase mode and the aminopropyl in hydrophilic interaction mode due to their complementarity. We expect the approach developed in this paper to be easily applicable to other organic complex mixtures of interest for planetary sciences, such as meteorites soluble organic matter or laboratory analogues of cosmic ices.

On our sample, we were able to strictly identify seven isomers of the major compounds present in Titan's tholins, and to discard six others including HMT, based on three complementary chromatographic separation coupled to HRMS analysis. All the detected molecules bear nitrogen within their structure and most of them are aromatics. This supports the idea of a tholins formation pathways passing through PANH.

## Funding sources

The research presented in this paper was partially funded through the French Programme National de Planétologie (PNP). NC acknowledges the European Research Council for their financial support (ERC Starting Grant PRIMCHEM, grant agreement n°636829). TG acknowledges the NASA Postdoctoral Program at the Goddard Space Flight Center, administered by Oak Ridge Associated Universities.